\documentclass[prc,aps,showpacs,floatfix,nofootinbib,letterpaper,twocolumn]{revtex4-1}
\usepackage{epsfig} 
\usepackage{amssymb} 
\usepackage{amsmath} 
\usepackage{amsfonts} 
\usepackage{color, graphicx} 
\usepackage{bm}

\def\lb{\langle} 
\def\rb{\rangle} 
 
\def\be{\begin{equation}} 
\def\ee{\end{equation}}

\newcommand{\ket}[1]{\left | #1 \right >} 
 
\newcommand{\braket}[2]{\left< #1 | #2\right>} 
\newcommand{\abs}[1]{\left | #1 \right |} 
\newcommand{\avg}[1]{\lb #1 \rb}

\begin{document} 

\title{Statistical-model description of $\gamma$ decay from
  compound-nucleus resonances} \author{P. Fanto$^{1}$,
  Y. Alhassid$^{1}$, and H.~A. Weidenm\"uller$^{2}$ }
\affiliation{$^{1}$Center for Theoretical Physics, Sloane Physics Laboratory, Yale University, New Haven, Connecticut 06520, USA\\
  $^2$Max-Planck-Institut f\"ur Kernphysik, D-69029 Heidelberg,
  Germany} \date{\today}
 
\begin{abstract} 
  The statistical model of compound-nucleus reactions predicts that
  the fluctuations of the partial $\gamma$-decay widths for a
  compound-nucleus resonance are governed by the Porter-Thomas
  distribution (PTD), and that consequently the distribution of total
  $\gamma$-decay widths is very narrow.  However, a recent
  experiment [Koehler, Larsen, Guttormsen, Siem, and Guber, Phys. Rev. C {\bf 88}, 041305(R) (2013)] reported large fluctuations of the
  total $\gamma$-decay widths in the $^{95}$Mo$(n,\gamma)^{96}$Mo*
  reaction, contrary to this expectation. Furthermore, in recent
  theoretical works it was argued that sufficiently strong channel
  couplings can cause deviations of the partial width distributions
  from PTD.   Here, we investigate
  whether the combined influence of a large number of nonequivalent
  $\gamma$-decay channels, each of which couples weakly to the
  compound-nucleus resonances, can modify the statistics of the
  partial widths. We study this effect in neutron
  scattering off $^{95}$Mo within a random-matrix model that includes
  coupling to the entrance neutron channel and to the large number of $\gamma$
  channels. Using realistic coupling parameters obtained from
  empirical models for the level density and the $\gamma$ strength
  function, we find that the PTD describes well the distribution of
  partial widths for all decay channels, in agreement with the
  statistical-model expectation. Furthermore, we find that the width
  of the distribution of the total $\gamma$-decay widths is insensitive
  to wide variations in the parameters of the $\gamma$ strength
  function, as well as to deviations of the partial-width
  distributions from the PTD. Our results rule out an explanation of
  the recent experimental data within a statistical-model description
  of the compound nucleus.
\end{abstract} 
 
 
\maketitle 

\section{Introduction}\label{intro}

It is widely accepted that low-energy neutron resonance scattering
from medium-mass and heavy nuclei is well described by the statistical
model~\cite{Mitchell2010}, in which the compound-nucleus (CN)
resonances are described as eigenstates of a Hamiltonian drawn from
the Gaussian orthogonal ensemble (GOE) of random-matrix theory.  The
statistical model predicts that the distribution of the partial widths
for each individual reaction channel follows the Porter-Thomas
distribution (PTD), i.e., a $\chi^2$ distribution in one degree of
freedom.  As a result, the total $\gamma$-decay width distribution is expected to be very narrow, resembling a $\chi^2$
distribution in many degrees of freedom.

In recent years, however, some experimental evidence was presented for
possible violations of the statistical-model predictions. The
distribution of neutron resonance widths obtained from $s$-wave
neutron scattering off Pt isotopes was found to be significantly
broader than the PTD~\cite{Koehler2010}. Moreover, a recent analysis
of the Nuclear Data Ensemble found a statistically significant
deviation of the distribution of neutron-resonance widths from the
PTD~\cite{Koehler2011}. There have been attempts to explain these
findings through non-statistical effects that emerge within
the statistical model~\cite{Celardo2011,Fyodorov2015,Volya2015}.
Other explanations focused on the analysis of the
data~\cite{Weidenmuller2010,Shriner2012}. For the case of Pt isotopes,
nearly all of the former explanations were ruled out by recent
work~\cite{Bogomolny2017,Fanto2017}.

Almost all of the experimental and theoretical works were focused on the
neutron resonance widths. However, a recent experiment that measured
total $\gamma$-decay widths from $s$- and $p$-wave neutron resonances in
neutron scattering off $^{95}$Mo found the total $\gamma$-decay width
distribution for each spin-parity class of resonances in $^{96}$Mo to be
significantly broader than the statistical model
predictions~\cite{Koehler2013}. It is important to understand whether
such large fluctuations are possible within the framework of the statistical
model.

It is known that the nonequivalence (i.e., different coupling strengths) of a set of channels corresponding to a particular CN decay mode leads to a reduction in the effective number of degrees of freedom describing the
set and thus to an increase in the total width
fluctuations~\cite{Drodz2000}. Individual $\gamma$ channels, each of
which is defined by the multipolarity of the transition and the energy
of the emitted $\gamma$ ray, are nonequivalent. However, in simulations
that were carried out in the experimental analysis, the nonequivalence
of the $\gamma$ channels was accounted for through the use of the $\gamma$
strength functions ($\gamma$SF) and of models for the level density.
These describe, respectively, the average partial width for a
transition of given multipolarity and $\gamma$-ray energy and the number
of accessible final states for the $\gamma$ decay. Thus, the
nonequivalence of the $\gamma$ channels cannot account for the disagreement between the measured and simulated total $\gamma$-decay width distributions of Ref.~\cite{Koehler2013}.

Recent theoretical works have shown that the violation of the
orthogonal invariance of the GOE due to the coupling to reaction
channels can lead to deviations of the partial width distribution from
the PTD~\cite{Celardo2011, Fyodorov2015, Volya2015}.
However, these works did not include a realistic description of the
$\gamma$-decay channels.
  In medium-weight and heavy nuclei, the number of
  $\gamma$-decay channels is very large. To avoid the explicit modeling of such a large number of channels, $\gamma$ decay is often modeled by a constant imaginary
  contribution to the effective Hamiltonian, but this model assumes that the total $\gamma$-decay width distribution is very narrow.
  Alternatively, using a realistic
  description of the coupling strength of the $\gamma$ channels, the
  width distribution for any number of $\gamma$ channels may be
  calculated using the analytic results of
  Ref.~\cite{Fyodorov2015}. However, given the large number of
  channels, the evaluation of formulas derived in Ref.~\cite{Fyodorov2015}
  is impractical. Therefore, it is not known yet whether the combined
  effect of a large number of weakly coupled $\gamma$-decay channels in
  the $^{95}$Mo$(n,\gamma)^{96}$Mo* reaction might modify the
  distribution of the partial $\gamma$-decay widths.

Furthermore, the total $\gamma$-decay width distribution depends not only on the partial $\gamma$-decay width fluctuations 
but also on the level density and $\gamma$SF.
Simulations of the distributions of total $\gamma$-decay widths for a
given spin-parity class of resonances use empirical formulas for the
$\gamma$SF and for the level density. In
Ref.~\cite{Koehler2013}, several different $\gamma$SF models were
used to generate statistical-model results, but the systematic
dependence of the simulated
distributions on the model parameters was not studied.
It is important to understand this dependence in order to know how sensitive the total
$\gamma$-decay width distributions are to the underlying partial width fluctuations.

Here, we investigate the role of the $\gamma$-decay channels in the
statistical model. First, we study the effect of the $\gamma$ channels on
the fluctuations of the partial widths. To facilitate the numerical
simulations, we group $\gamma$ channels of the same multipolarity that are
close in energy into a single `representative' channel. We expect
that such coarse-graining of the channels does not change the
qualitative results if a sufficient number of representative channels
is used. We use empirical parameterizations of the level density,
$\gamma$SF, and neutron strength function to determine the average
channel couplings. We then calculate the distributions of the partial
neutron widths and of the partial $\gamma$-decay widths by using a large
number of GOE realizations of the CN Hamiltonian.
We find no deviation from the PTD and thus confirm the traditional
expectation of the statistical model.

Next, we address the distribution of total $\gamma$-decay widths,
focusing attention on the widths and peak locations of these
distributions. We systematically vary the parameters of the
$\gamma$SF, assuming the partial widths are described by PTD. We find
virtually no change in the width of the total $\gamma$-decay width
distribution for a broad range of the $\gamma$SF
parameters. Furthermore, although the peak location may be reproduced
for any individual spin-parity class of resonances by parameter
adjustments, we cannot obtain agreement of the peak locations
with experiment for all spin-parity classes through such adjustments. This result indicates a
serious shortcoming of the empirical $\gamma$SF expressions for the
$^{96}$Mo compound nucleus.

Finally, we investigate whether the total $\gamma$-decay width
distribution is sensitive to deviations in the partial $\gamma$-decay
width distribution from PTD, which can occur for sufficiently strong
coupling of the neutron channel. We find that these modified
fluctuations in the partial widths have virtually no effect on the
total $\gamma$-decay width distribution.

In the following, we comment on a limitation of our study. Our
goal is to investigate whether the experimental
findings of Ref.~\cite{Koehler2013} can be explained within the GOE statistical model
of the CN as defined in Ref.~\cite{Mitchell2010}.  Although the GOE statistical model forms the basis of the statistical
theory of CN reactions and is widely used in applications, the GOE effectively assumes random $n$-body interactions, where $n$ is the number of particles.
From a physical point of view, it would be preferable to use more realistic
statistical models for the CN that account for the predominantly
two-body character of the residual nuclear interaction.
Such models are the $k$-body embedded ensemble
EGOE($k$) with $k=2$~\cite{Mon1975} and the two-body random ensemble (TBRE)~\cite{French1970}; see Refs.~\cite{Kota2001,Benet2003,Gomez2011} for reviews. 
In contrast to the GOE, however, the spectral fluctuation properties of these models cannot be
determined analytically.  The limited information that exists
has been gained numerically~\cite{Kaplan2000,Kota2001,Celardo2007,Volya2011}. 
In Ref.~\cite{Kota2001}, the distribution of transition strengths for EGOE(2) was found to be well
described by the PTD.\footnote{The residual interaction in nuclei is sufficiently strong for EGOE(2) to be a good approximation to the more realistic mixed ensemble EGOE(1) + EGOE(2).}  Some evidence exists of PTD violation in
the TBRE~\cite{Kaplan2000,Volya2011}, but this violation mainly occurs in the tails of its spectrum~\cite{Kaplan2000} and 
is ascribed to a lack of complete mixing of the basis states. 
That interpretation is consistent with results of shell-model
calculations for $sd$-shell nuclei~\cite{Brown1984,Zelevinsky1996},
which showed that the shell-model eigenvector components were practically Gaussian (non-Gaussian)
in the center (tails) of the spectrum.\footnote{A shell-model Hamiltonian can be considered as one particular realization of EGOE(1) + EGOE(2).}
In actual nuclei, such tails would
comprise the ground state and a number of low-lying excited states, 
whereas the CN resonances we consider in this work are far from these tails. Shell model calculations of electromagnetic transition strength distributions in $sd$-shell~\cite{Adams1997} and in $pf$-shell~\cite{Hamoudi2002} nuclei were also found to follow the PTD.
In view of these facts, using the EGOE($2$) or TBRE for CN calculations does
not seem to be a pressing need.  
Moreover, and most importantly, the GOE
model for the CN yields analytical results, which are very useful for numerical
implementation in applied codes, whereas the EGOE($2$) and TBRE 
have so far not provided such analytical expressions.  
Therefore, we limit our studies here to the GOE description of the CN.

The outline of this article is as follows. In Sec.~\ref{model}, we present
our model for studying the statistics of the partial widths with the
large number of $\gamma$-decay channels taken into account. In Sec.~\ref{app}, we
discuss the physical parameters used to apply this model to the $n +
^{95}$Mo reaction. In Sec.~\ref{partial}, we show that the PTD
provides an excellent description of the partial width statistics for
the reaction considered. In Sec.~\ref{vary}, we show the effect of
varying the $\gamma$SF parameters on the simulated total $\gamma$-decay
width distribution. In Sec.~\ref{beyond-ptd}, we study the effect of
modified partial $\gamma$-decay width distributions on the total $\gamma$-decay width distribution. Finally, in Sec.~\ref{conclusion}, we
summarize our results.

\section{Statistical model of CN resonances}\label{model}

In the absence of direct reactions, the scattering matrix ($S$ matrix)
for  CN reactions is given by
\be\label{S1}
S_{cc^\prime}(E)  = \delta_{cc^\prime} - 2\pi i \sum_{\mu \nu} W_{\mu c} \left( E - H^{\rm eff}\right)^{-1}_{\mu \nu} W_{\nu c} \;,
\ee where $c,c^\prime$ denote reaction channels, and $\mu,\nu$ denote the internal CN states.  Eq.~\eqref{S1} 
depends on the effective non-Hermitean Hamiltonian $H^{\rm eff}$ that
governs the CN resonances and is given by~\cite{Mitchell2010}
\be\label{heff-1}
\begin{split}
H^{\rm eff}_{\mu\nu} = H^{\rm GOE}_{\mu\nu} + & \sum_c \mathcal{P} \int dE^\prime \frac{W_{\mu c}(E^\prime) W_{\nu c}(E^\prime)}{E - E^\prime} 
\\ &  - i\pi \sum_c W_{\mu c}(E) W_{\nu c}(E) \;.
\end{split}
\ee
Here $H^{\rm GOE}$ is a GOE random matrix, $W_{\mu c}(E)$ denotes the coupling of the
 state $\mu$ of a fixed basis of the internal state space to the
channel $c$ at the incident neutron energy $E$, and $\mathcal{P}$
is the principal-value integral. The coupling constants $W_{\mu
  c}(E)$ form an $N \times \Lambda$ real matrix $W(E)$, where $N$ is
the dimension of the internal space of CN resonances and $\Lambda$ is the
number of open channels.

Ignoring direct reactions, the coupling matrix $W$ in the
basis of physical channels $c$ satisfies~\cite{Mitchell2010} 
\be
 (W^T W)_{cc^\prime} = \delta_{cc^\prime} \kappa_c \lambda/\pi \label{a} \;,
  \ee
where $\kappa_c$ are dimensionless parameters determining the strength
of the coupling and $\lambda = ND/\pi$ is the GOE energy scale
parameter with $D$ being the average spacing of GOE eigenstates in the
middle of the spectrum~\cite{Mitchell2010}.  We choose $c=1$ to be the
neutron channel and $c > 1$ to be the $\gamma$ channels.

According to Eq.~(\ref{a}), the $\Lambda$ vectors $\vec
W_c/\sqrt{\kappa_c \lambda/\pi}$ ($c = 1,...,\Lambda)$ are
orthonormal. We choose these and additional $N-\Lambda$ orthonormal
vectors that are orthogonal to them as a basis for the CN states. The GOE is invariant under
such an orthogonal transformation. In this basis, the effective
Hamiltonian takes its canonical form
 \be\label{heff-canonical} \tilde
H^{\rm eff}_{\mu\nu} = H^{\rm GOE}_{\mu \nu} + \delta_{\mu \nu} V_\mu \;,
\ee 
where the first $\Lambda$ elements of the diagonal term on the
r.h.s.~are
 \be\label{Vc} V_c = \lambda\left(\frac{1}{\pi} \mathcal{P}
  \int_0^\infty dE^\prime \frac{\kappa_c}{E-E^\prime} -
  i\kappa_c\right) \ee
  for $c = 1,...,\Lambda$, and $V_\mu = 0$ for $\mu > \Lambda$.

The principal-value integral in (\ref{Vc}) describes a
real diagonal shift to the GOE Hamiltonian. For the reasons explained in Sec.~\ref{pval}, we
neglect it for the neutron channel and all the $\gamma$ channels. Consequently,
the non-statistical diagonal shifts to the GOE Hamiltonian in
Eq.~(\ref{heff-canonical}) become purely imaginary
 \be V_c = -i\kappa_c \lambda\;. \ee

\subsection{Partial widths}

The effective Hamiltonian in Eq.~(\ref{heff-canonical}) provides the most convenient way to study
partial widths of the CN resonances to decay into individual channels within the framework of the
statistical model. We consider the limit of isolated
resonances.  The resonance energies and widths are determined, respectively, by the
real and imaginary parts of the eigenvalues of $H^{\rm eff}$.

Rewriting the $S$ matrix in Eq.~(\ref{S1}) in the basis used in
Eq.~(\ref{heff-canonical}), we obtain
 \be\label{S2} S_{cc^\prime} =
\delta_{cc^\prime} - 2i \lambda \left(\kappa_c
  \kappa_{c^\prime} \right)^{1/2} \left(E - \tilde H^{\rm
    eff}\right)^{-1}_{cc^\prime}\;.
    \ee 
 The matrix $\tilde H^{\rm eff}$ is a complex symmetric matrix and can be diagonalized by a complex orthogonal transformation
$U$, yielding 
\be\label{heff-diag}
\left(U^T \tilde H^{\rm eff} U\right)_{\mu \nu} =
\delta_{\mu \nu}\left(E_\mu - i \frac{\Gamma_\mu}{2}\right) \;.  \ee
Under this transformation, the diagonal $S$-matrix element $S_{cc}$
becomes 
\be
 S_{cc} = 1 - 2i \lambda \sum_{\mu} \frac{\kappa_c
  U_{c\mu}^2}{E - E_\mu + i(\Gamma_\mu/2)} \;. 
 \ee 
   The partial width $\Gamma_{\mu c}$ for resonance $\mu$ ($\mu = 1,...,N$) to decay into
channel $c$ is thus given by
\be\label{partialwidth} 
\Gamma_{\mu c} = \frac{2 \lambda \kappa_c |U_{c \mu}|^2 }{(U^T U^*)_{\mu\mu}}\;,
\ee
where the term $(U^TU^*)_{\mu\mu}= \sum_\nu |U_{\nu\mu}|^2$ in the denominator is the squared norm of the column vector $\mu$ of the matrix $U$, known as the Petermann factor \cite{Mitchell2010,Frahm2000}  The inclusion of the Petermann factor in the definition of the partial width ensures that the sum of all partial widths is equal to the total resonance width, i.e., $\sum_c \Gamma_{\mu c}=\Gamma_\mu$ \cite{Moldauer1964}.  In the limit of isolated resonances, $U$ is a real matrix and  $(U^T U^*)_{\mu\mu}=1$.
Once the values of the coupling parameters
$\kappa_c$ are specified, one can diagonalize a large number of
realizations of the effective Hamiltonian and determine the partial
width distributions for the various channels.

In principle, the coupling parameters $\kappa_c$ are determined from
the average $S$ matrix 
\be
\avg{S}_{cc^\prime} =\delta_{cc^\prime} \frac{1-\kappa_c}{1+\kappa_c} \;.
\ee
 This requires a realistic optical-model
calculation. Instead, we determine approximate values $\kappa_c$ from
the partial widths obtained in first-order perturbation theory. In
this case, $U_{c \mu}$ are the elements of GOE eigenvectors and as
such are independent Gaussian random variables with zero mean and variance
of $1/N$.  Taking the GOE average of (\ref{partialwidth}), we find
\be\label{coupling}
 \kappa_c = \frac{\pi}{2} \frac{\avg{ \Gamma_{\mu  c}}}{D_{J^\pi}} \;,
 \ee 
 where $\avg{ \Gamma_{\mu c}}$ is the
average partial width to decay into channel $c$, and $D_{J^\pi}$ is the
average spacing of CN resonances with spin-parity $J^\pi$.
The average partial widths can be estimated using empirical
parameterizations of the strength functions for the neutron and $\gamma$
channels, and of the level density.

\subsection{Representative $\gamma$ channels}\label{rep}

Each $\gamma$-decay channel $f$ is specified by the multipolarity and type
(i.e., electric or magnetic) of the emitted $\gamma$ ray and by the final
state (energy $E_f$ and spin-parity values $J_f^\pi$).
The number of final states to which each resonance may decay is
governed by the level density $\rho(E_f, J^\pi_f)$.
The average partial width to decay from a resonance $\mu$ of energy
$E_\mu$ and spin-parity $J^\pi$ to a channel $f$, divided by the
average resonance spacing, is given by \be\label{gammapartial}
\frac{\avg{ \Gamma^{J^\pi}_{\gamma\mu fXL}}}{D_{J^\pi}} =
E_\gamma^{2L+1} f_{XL}(E_\gamma) \;,
\ee 
where $E_\gamma = E_\mu - E_f$ is
the energy of the emitted $\gamma$ ray;
$XL$ specifies the type and multipolarity of the transition; and
$f_{XL}(E_\gamma)$ is the corresponding 
$\gamma$SF. The average total $\gamma$-decay width $\avg{ \Gamma^{J^\pi}_{\gamma\mu}}$ of resonance $\mu$ is obtained by
summing (\ref{gammapartial}) over the allowed final states
\be\label{total-width-integral}
\begin{split}
& \avg{ \Gamma^{J^\pi}_{\gamma\mu}} = \sum_{XL}\sum_f \avg{ \Gamma^{J^\pi}_{\gamma\mu fXL}} \\  & = D_{J^\pi}\sum_{XL} \int_0^{E_\mu}d E_\gamma E_\gamma^{2L+1} f_{XL}(E_\gamma)  \sum_{J^\pi_f} \rho(E_\mu-E_\gamma,J^\pi_f)\,.
\end{split}
\ee  

Here we consider only dipole transitions $L=1$ (both electric and magnetic) as these give
the main contributions to the total width. As mentioned above, because of the large density
of final states, it is impractical to include all
of the final states accessible by dipole $\gamma$-ray emission.
Instead, in our model each representative $\gamma$-decay channel $c$
describes a group of physical $\gamma$ channels $f$ that are close in
final energy $E_f$. In practice, we generate a set of representative
final levels whose average density is proportional to the actual level
density.
We set $\kappa_c$ for each representative channel $c$ to be
\be\label{repcoup}
 \kappa_c = \frac{\pi}{2}\sum_{f \in c} \frac{\avg{
    \Gamma^{J^\pi}_{\gamma\mu f XL}}}{D_{J^\pi}} \;,  
    \ee 
which is obtained from Eq.~(\ref{coupling}) by summing over all physical
channels $f$ in $c$.
We choose the summation in Eq.~(\ref{repcoup}) such that the density
of representative channels $c$ is related to the density of physical
channels $f$ by an energy-independent constant $G = (\Lambda-1)/\Lambda_{\gamma f}$, where $\Lambda-1$ is the total
number of representative $\gamma$ channels in our model and $\Lambda_{\gamma f}$ is
the total number of physical $\gamma$ channels. Finally, we normalize the
coupling parameters $\kappa_c$ to satisfy
 \be\label{repnorm}
  \sum_{c =  2}^\Lambda \kappa_c = \frac{\pi}{2} \frac{\avg{
    \Gamma^{J^\pi}_{\gamma \mu;\;\rm exp}}}{D_{J^\pi}} \;,
 \ee where
$\avg{\Gamma^{J^\pi}_{\gamma \mu;\;\rm exp}}$ is the average total width
determined from the experiment~\cite{Koehler2013}.

A proper method of coarse graining should yield the same physical
results at any scale. Our method does not guarantee this; for
sufficiently small $\Lambda$, our model could yield effects that
vanish as $\Lambda$ is increased. However, we claim that, for large
enough $\Lambda$, the model results will be qualitatively the same as
the physical results. Our argument is as follows. Below some coupling
strength, each individual channel may be treated perturbatively. All
physical $\gamma$ channels lie below this bound. As discussed above, no
single $\gamma$ channel is strong enough to perturb the GOE dynamics. If
we choose $\Lambda$ such that the strongest representative $\gamma$-decay
channel may be treated perturbatively, then the qualitative behavior
caused by the set of representative $\gamma$ channels should be similar
to the physical case.

\subsection{Principal-value integral}\label{pval}

The principal-value integral on the r.h.s.~of
Eq.~(\ref{heff-1}), also known as the Thomas-Ehrman shift, contributes
a real non-statistical term to the effective Hamiltonian and thus
appears to be a possible source of deviations from GOE statistics.  In
Ref.~\cite{Volya2015}, the real shift due to the neutron channel was
proposed as a possible explanation of the deviation from the PTD
observed in Ref.~\cite{Koehler2010}. Assuming an energy-independent
coupling, it was shown that a real shift in the single-channel case
leads to an energy dependence of the average partial width on the
scale of the entire spectrum but locally the fluctuations are still
described by the PTD~\cite{Bogomolny2017}. Recent work showed that the
real shift does not affect the PTD of the normalized widths even when
a realistic energy dependence of the couplings is
included~\cite{Fanto2017}.
 In the calculations that follow, we thus ignore the real shifts in all channels. 

\section{Application to $n + ^{95}$Mo}\label{app}
 
Here we study the $^{95}$Mo$(n,\gamma)^{96}$Mo* reaction using the
statistical model discussed in Sec.~\ref{model}. The ground
state of $^{95}$Mo has spin-parity $5/2^+$. Therefore, the CN
resonances in $^{96}$Mo* have spin-parity of $J^\pi=2^+, 3^+$
for $s$-wave neutrons and $J^\pi=1^-, 2^-, 3^-, 4^-$ for $p$-wave
neutrons. We study each of these cases.

\subsection{Level density}

Within our model, the calculation of the statistics of $\gamma$-decay
widths requires realistic parameterizations of the level density and
the $\gamma$SF of the compound nucleus $^{96}$Mo. For the
level density, we use the back-shifted Fermi gas
formula~\cite{Huizenga1972}, also known as the back-shifted Bethe Formula (BBF),
together with the spin-cutoff model~\cite{Ericson1960} and the
assumption of equal densities for both parities.  We have
\be\label{LD}
 \rho(E,J^\pi) =
 f(J) \frac{\sqrt{\pi}}{24 \, a^{1/4}}\frac{e^{2\sqrt{a(E-\Delta)}} }{(E-\Delta)^{5/4}} \;, 
 \ee
where $a$ and $\Delta$ are, respectively, the single-particle level density and
backshift parameters, and $f(J)= \rho(E,J)/\rho(E)$ is the spin distribution
\be \label{spin}
f(J) = {(2J+1) \over 2\sqrt{2 \pi} \sigma_c^3} e^{-{J(J+1) \over 2 \sigma_c^2}}\,.
 \ee
The parameter $\sigma_c$ in (\ref{spin}) is known as the spin-cutoff parameter, for which we use ~\cite{Krticka2008,Sheets2009} 
\be \sigma_c^2 =
0.0888A^{2/3}\sqrt{a(E-\Delta)}
 \ee 
with $A$ being the mass number. The values for $a$ and $\Delta$, determined by fitting the BBF to
level counting data at low energies and the neutron resonance data at
the neutron threshold energy~\cite{Ozen2019}, are given in
Table~\ref{T1}.

\begin{table}[h!]
\caption{\label{T1} Parameters for the level density~\cite{Ozen2019} and $\gamma$SF~\cite{Krticka2008} in $^{96}$Mo (see text).  
}
\begin{tabular}{ l c c c c  }
\hline\hline
a (MeV$^{-1}$)& $\Delta$ (MeV) & $E_{\rm G}$ (MeV)  & $\Gamma_{\rm G}$ (MeV) & $\sigma_G$ (mb) \\   \hline 
11.41 & 0.85 & 16.2 & 6.01 & 185.0 \\ 
 $\Delta_{\rm G}$ (MeV) & $E_{\rm SF}$ (MeV) & $\Gamma_{\rm SF}$ (MeV) & $\sigma_{\rm SF}$ (mb) & C (MeV$^{-12}$) \\ \hline 
2.55 & 8.95 & 4.0 & 0.4 & 1.0 \\ \hline\hline
\end{tabular}

\end{table}

We use the level density to generate a spectrum of final states, which
is necessary to calculate the average partial widths for the $\gamma$
transitions [see Eq.~(\ref{gammapartial})].  To generate these final
states, we follow a similar procedure to that used to produce each realization in the DICEBOX code~\cite{Becvar1998}.  Below a threshold
energy $E_{\rm th} = 2.79$ MeV, we include a complete set of experimentally measured discrete levels~\cite{Krticka-priv}.  Above $E_{\rm th}$, we draw energies that follow the corresponding level density.  The total number $ N_{J_f^\pi}$ of
final energies we draw for spin-parity class $J_f^\pi$ is
given by
 \be\label{nfinal} 
 N_{J_f^\pi} = \int_{E_{\rm th}}^{S_n} \rho(E,J_f^\pi) dE \;, 
\ee 
where the allowed final spins and parities $J_f^\pi$ are determined by the selection rules for
$E1$ and $M1$ transitions, and $S_n$ is the neutron separation energy.  In
contrast to the DICEBOX approach, we do not average over realizations.
Instead, we use only one fixed set of final energies for each
spin-parity class, neglecting the fluctuations of the final states.
To create the representative channels described in Sec.~\ref{rep} from
these final states, we calculate the average partial width $\avg{
  \Gamma^{J^\pi}_{\gamma\mu f XL}}$ for each of these final energies.
We then collect the final energies into groups, each of which consists
of the same number of neighboring final energies.  This group
corresponds to a representative channel $c$.  We calculate the
parameters $\kappa_c$ by using Eq.~(\ref{repcoup}) for each group.

\subsection{$\gamma$SF}\label{gsf}

\begin{figure}[bth]
\includegraphics[width=0.45\textwidth]{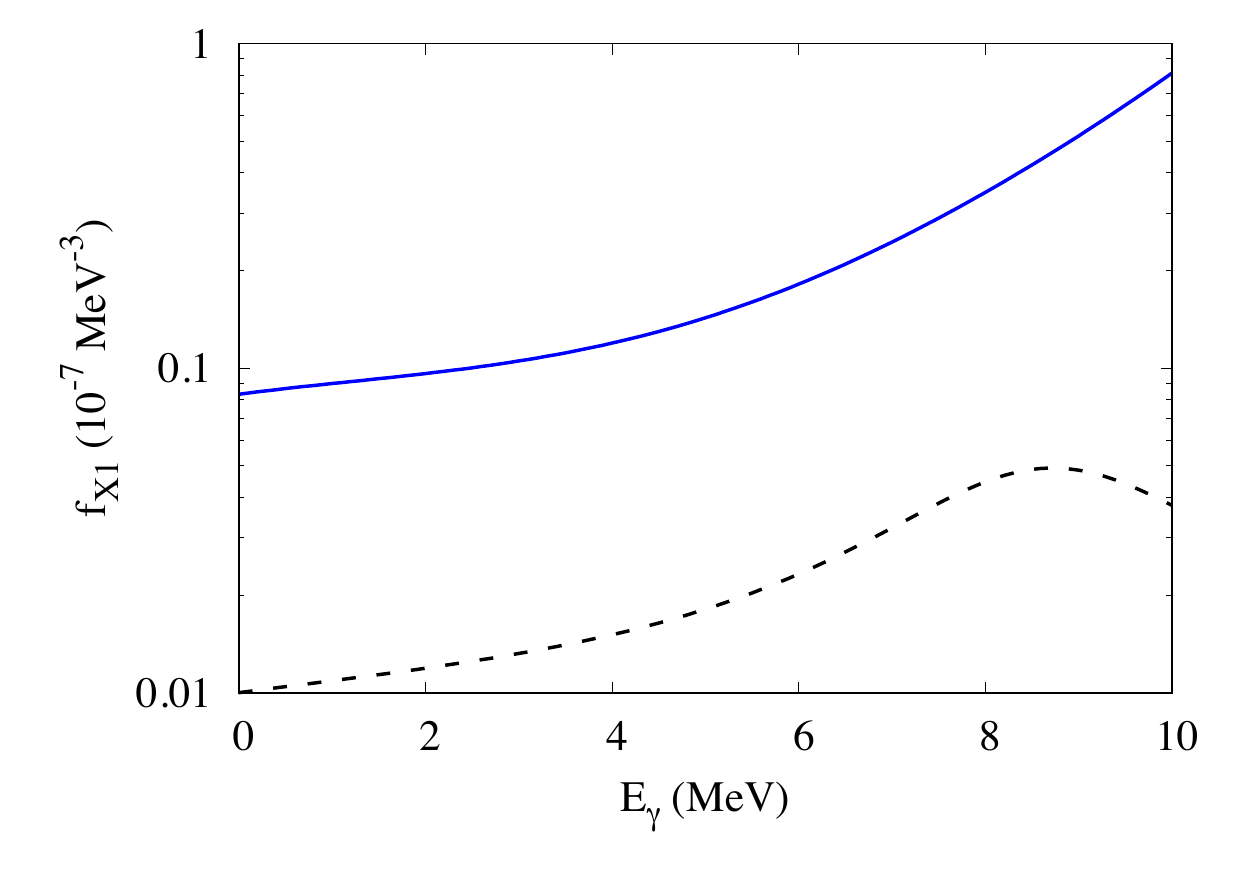}
\caption{\label{ldgsf}
$\gamma$ strength functions vs.~$\gamma$-ray energy $E_\gamma$ used in our
  calculations: $E1$ (solid blue line) and $M1$ (dashed black line).}
\end{figure}

We use the $E1$ and $M1$ $\gamma$SF of Refs.~\cite{Krticka2008,Sheets2009}.  
For the $E1$ $\gamma$SF, we use the generalized Lorentzian (GLO) model given in Eq.~(6) of
Ref.~\cite{Sheets2009} 
\begin{eqnarray}\label{E1}
f_{E1}(E_\gamma) = \frac{\sigma_{\rm G} \Gamma_{\rm G}}{3 (\pi \hbar c)^2}&\bigg[ \frac{ E_\gamma \Gamma(E_\gamma, T)}{(E_\gamma^2-E_{\rm G}^2)^2 + E_\gamma^2\Gamma(E_\gamma,T)^2} 
\nonumber \\& + 0.7 \frac{4\pi^2 \Gamma_{\rm G} T^2}{E_{\rm G}^5}\bigg] \;.
\end{eqnarray} 
Here $T$ is a temperature parameter given by $T^2 = (S_n - E_\gamma
- \Delta_{\rm G})/a$, $S_n = 9.154$ MeV is the neutron separation
energy in $^{96}$Mo, and 
\be 
\Gamma(E_\gamma,T) = \Gamma_G
\frac{E_\gamma^2 + 4\pi^2 T^2}{E_G^2} \;. 
\ee 
The M1 strength function is given by
\be\label{M1}
 f_{M1}(E_\gamma) = \frac{1}{3(\pi\hbar c)^2}
\frac{\sigma_{\rm SF}E_\gamma \Gamma_{\rm SF}^2}{(E_\gamma^2 - E_{\rm
    SF}^2)^2 + E_\gamma^2 \Gamma_{\rm SF}^2} + C \;. \ee
The Lorentzian term on the r.h.s.~of Eq.~(\ref{M1}) describes the
spin-flip term [see Eq.~(5) of Ref.~\cite{Sheets2009}], and the constant $C$ is the single-particle term.  The
values of the parameters in Eqs.~(\ref{E1}) and (\ref{M1}) are given
in Table~\ref{T1}.  In Fig.~\ref{ldgsf} we show the $E1$ and $M1$ $\gamma$SF of Eqs.~\eqref{E1} and \eqref{M1}, respectively,  for $^{96}$Mo*.  

We determine the coupling $\kappa_n$ in the neutron channel from the
neutron strength function.  For $s$-wave resonances, we take the
neutron strength function parameter $S_0 = 0.47 \times 10^{-4}$
eV$^{-1/2}$ from the RIPL-3 database~\cite{ripl}.  The average partial
width for these resonances is then given by $\langle \Gamma_{\mu n} \rangle/ D_{J^\pi} =
S_0 \sqrt{E}$ where $E$ is the energy of the incoming neutron.  We ignore the energy dependence of the
average neutron width, and take its value for $E = 10$ eV, which
is at the highest end of the experimental range of
Ref.~\cite{Koehler2013} (see Fig. 5 of this reference).  The coupling constant $\kappa_n$ is then determined from \eqref{coupling}.  For
simplicity, we use for $p$-wave resonances the same coupling as for the $s$-wave resonances (see
Sec.~\ref{partial}).

\section{Partial width distributions}\label{partial}

In our simulations, we used 100 realizations of a GOE matrix of dimension $N = 1000$ and
$\Lambda = 401$ channels.  These channels consist of one neutron
channel, 200 $E1$ representative channels, and 200 $M1$ representative
channels. For each GOE realization, we diagonalized the Hamiltonian in Eq.~\eqref{heff-canonical} to determine its eigenstates, which compose the columns of the matrix $U$ in \eqref{heff-diag}. We took the eigenstates $\mu$ from the middle half of the spectrum to avoid unphysical effects due to the finite bandwidth of the GOE
matrices.  According to Eq.~\eqref{partialwidth}, the partial width $\Gamma_{\mu c}$ of resonance $\mu$ to decay into channel $c$ is proportional to $|U_{c\mu}|^2/(U^T U^*)_{\mu\mu}$. This term is equivalent to the projection of the normalized complex eigenvector $| \mu \rangle$ of the effective Hamiltonian onto the channel vector $| c \rangle $, i.e. $|\langle c | \mu \rangle |^2 = |U_{c\mu}|^2/(U^T U^*)_{\mu\mu}$.  We define  
\be\label{sq-proj} 
g_{\mu c} = \abs{\braket{c}{\mu}}^2/\overline{\abs{\braket{c}{\mu}}^2} \;,
\ee 
where the bar indicates the average value of the entire data set.  According to  Eq.~(\ref{partialwidth}), the fluctuations of the partial widths $\Gamma_{\mu c}$ are determined by the fluctuations of $g_{\mu c}$.
In the following, we will refer to the normalized squared projections $g_{\mu c}$ simply as the widths. 
 We study both the energy-dependent average widths $\avg{g_{\mu c}}$ and the
fluctuations of the reduced widths $\hat{g}_{\mu c} = g_{\mu
  c}/\avg{g_{\mu c}}$.  If the couplings to the channels do not significantly
perturb the GOE behavior of the resonances, then the average squared
projection for any channel in our model will be constant,
i.e., independent of the real resonance energy.  Moreover, the
fluctuations of the squared projections will follow the PTD.

\begin{figure}[b]
\includegraphics[width=0.45\textwidth]{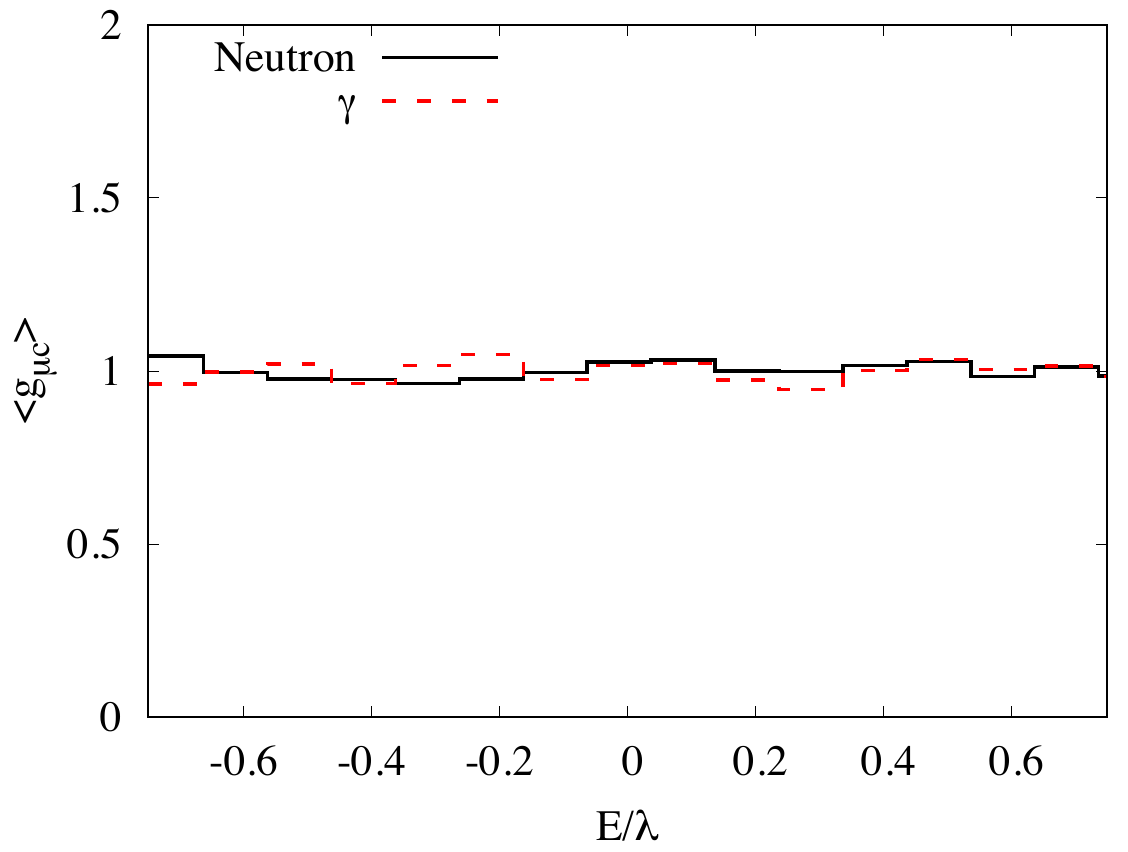}
\caption{\label{redavg20} Average widths $\avg{g_{\mu c}}$
  from Eq.~(\ref{sq-proj}) for the neutron channel (black solid line)
  and the strongest $\gamma$ channel (red dashed line) as a function of
  the real part of the resonance energy.  All energies are in units of
  the GOE parameter $\lambda$ (see Sec.~\ref{model}).}
\end{figure}

\begin{figure}[h!]
\includegraphics[width=0.48\textwidth]{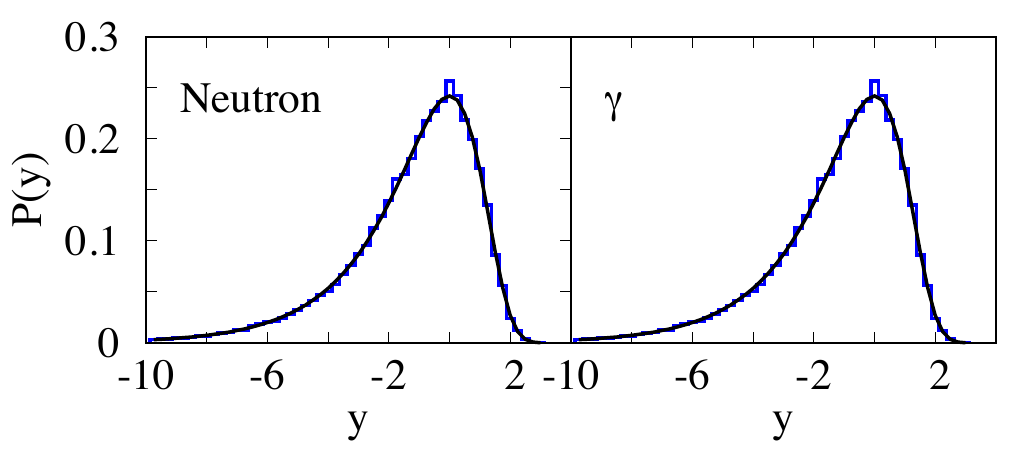}
\caption{\label{dist20} Distribution of $y = \ln x$, where $x = \hat
  g/\avg{\hat g}$ is the normalized reduced partial width
  for the neutron channel (left panel) and most strongly coupled $\gamma$ channel (right panel) [see Eq.~\eqref{sq-proj}].  The solid black line
  is the PTD of Eq.~\eqref{PTD}.}
\end{figure}

In Fig.~\ref{redavg20}, we show the average partial width
$\avg{g_{\mu c}}$ for the neutron channel and the most strongly
coupled $\gamma$ channel for initial resonances with spin-parity $1^-$.  The
average width is a constant across the spectrum, in agreement with the
 GOE expectation.  The average widths are the same for the
neutron and $\gamma$ channels because of the normalization in
Eq.~(\ref{sq-proj}).  The histograms in Fig.~\ref{dist20} show the distribution
of $ y = \ln x$, where $x = \hat g/\avg{\hat{g}}$ for the neutron channel and
most strongly coupled $\gamma$ channel.  The PTD for $y$  (solid line)
\be\label{PTD}
\mathcal{P}(y) = \sqrt{\frac{x}{2\pi}} e^{-x/2} 
\ee
 is seen to be in excellent
agreement with the model calculations.  

We found similar results for other $\gamma$
channels (besides the most strongly coupled one), and for other spin-parity values of the initial resonances. 
These results, as well as the computer codes used for the
calculation, are provided in the Supplemental Material~\cite{supp}.  For the
$p$-wave resonances, we should in principle use a weaker coupling for
the neutron channel. However,  for simplicity we used the $s$-wave neutron
channel coupling.  Since we find no
deviation from the usual statistical behavior for this stronger
coupling, we conclude that there will be no deviation for the more
realistic $p$-wave coupling.

\section{Variation of the $\gamma$ strength function}\label{vary}

Statistical-model results were generated in Ref.~\cite{Koehler2013}
for various combinations of $\gamma$SF and level-density models. We
do not undertake a similarly thorough investigation here. Rather, we
intend to establish whether it is possible to reproduce either the
peak locations or the widths of the experimental total $\gamma$-decay width
distributions within large variations of the parameters of the
$\gamma$SF defined in Sec.~\ref{gsf}. We focus on the strength
function because it is less well determined than the level density.

Our method for generating a total $\gamma$-width distribution
is essentially the same as that of Ref.~\cite{Koehler2013} and follows
 the first step of the DICEBOX approach~\cite{Becvar1998}.
We use as input the $\gamma$SF parameters and a set of final states
with allowed values $J_f^\pi$ determined by the selection rules and
their corresponding level densities. We then calculate a total $\gamma$-decay width by
summing over the partial widths for transitions to each of the
final states $f$.  The partial widths for resonances of spin-parity
$J^\pi$ to decay with $\gamma$ radiation of multipolarity $XL$ are given
by 
\be\label{sim-partial} \Gamma^{J^\pi}_{\gamma \mu f XL} =
\avg{\Gamma^{J^\pi}_{\gamma \mu f XL}} x_f^2 \;,
\ee 
where $x_f$ is drawn
from a normal distribution with zero mean and unit variance.  $x_f^2$
is thus distributed according to the PTD.  This procedure is repeated
1000 times to obtain a set of total widths.

We vary the parameters $E_G$, $\Gamma_G$, and
$\sigma_G$ of the $E1$ $\gamma$SF (\ref{E1}) by factors of
2 in either direction to make them greater or smaller than their
values given in Table~\ref{T1}. These variations dramatically change
the strength of the $E1$ component of the $\gamma$SF, making the $M1$ component either more
or less significant relative to the $E1$ component. 

\begin{figure}[h]
\includegraphics[width=0.48\textwidth]{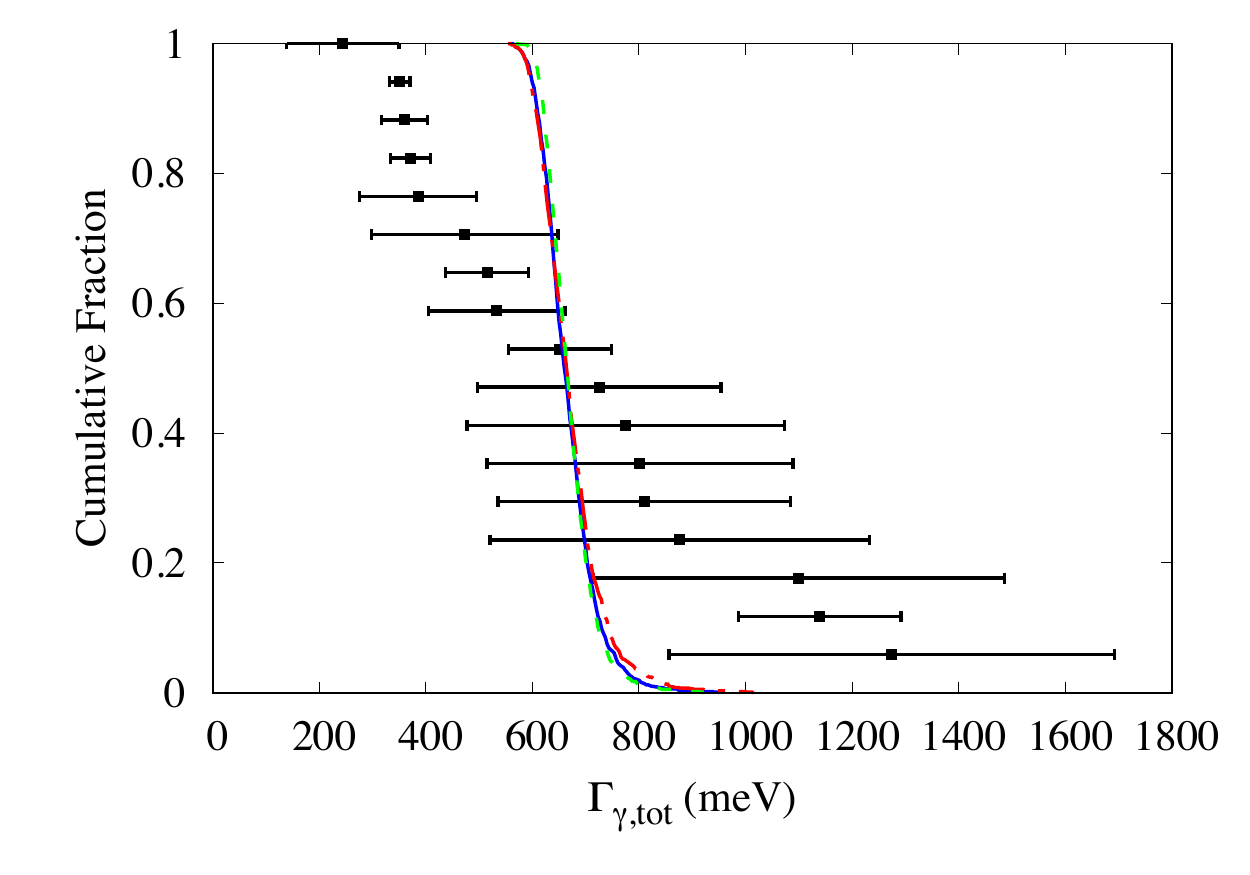}
\caption{\label{ggvar} The cumulative fraction of the total $\gamma$-decay widths for the $1^-$
  resonances. The black squares with error bars are
  experimental results from Ref.~\cite{Koehler2013}. The blue solid
  line is the simulation for the parameters in Table \ref{T1}.  The red
  dashed line is the result for $2\times\Gamma_G$ in Table \ref{T1}, and the
  green dashed-dotted line is the result for $(1/2)\times\Gamma_G$.
  The simulation results are normalized to fit the experimental
  average total width.}
\end{figure}

We find that these variations of the $\gamma$SF have no significant
effect on the widths of the total $\gamma$ decay width distributions. We show a representative
result in Fig.~\ref{ggvar} for the $1^-$ resonances. We plot the
cumulative fraction, i.e., the fraction of total widths greater than a
given width $\Gamma_{\gamma,\rm tot}$.
The simulated partial widths have been normalized such that their sum
 reproduces the average value of the experimentally observed total width.  This
normalization does not affect the relative contributions of the
various partial widths and thus does not change the width of the
distribution.  Our results, shown in Fig.~\ref{ggvar}, exhibit only weak dependence on the parameter $\Gamma_G$ and are compared with the experimental cumulative fraction measured in Ref.~\cite{Koehler2013}. We
conclude that the experimental distribution of the total $\gamma$-decay
width $\Gamma_{\gamma,\rm tot}$ is significantly broader than the theoretical distribution
obtained in the statistical model, and cannot be reproduced by a
reasonable variation of the $\gamma$SF parameters.

The average total $\gamma$ width, i.e., the peak of the total $\gamma$-decay width
distribution, is sensitive only to the level density and the
$\gamma$SF [see Eq.~(\ref{total-width-integral})]. In
Ref.~\cite{Koehler2013}, there were large discrepancies between the
simulated and experimental average total $\gamma$-decay widths. We find such
discrepancies for our choice of level density and $\gamma$SF as well.
In Table \ref{T2}, we list the average total widths calculated for our baseline
parameter values versus the experimental values for all spin-parity classes of resonance. 
The single-parameter variations in the $\gamma$SF we considered above
also influence the average total width. 
For any given spin-parity class of
resonances, we are able to reproduce the average width by varying one
parameter. For instance, for the $2^+$ resonances, multiplying the
parameter $\Gamma_G$ of Table \ref{T1} by a factor $f_G = 1.13$ brings the
average total width into excellent agreement with the experimental
value $\Gamma_{\rm tot}^{2^+} = 206$ meV. However, none of these
simple parameter adjustments reproduces simultaneously  the average total $\gamma$-decay widths for all the
spin-parity classes.

\begin{table}[h!]
\caption{\label{T2} Comparison of simulated average total $\gamma$-decay widths $\avg{\Gamma_{\gamma,\rm sim}}$ with the experimental widths $\avg{\Gamma_{\gamma,\rm exp}}$. The simulated results are calculated using the baseline parameter values for the strength functions.}
\begin{tabular}{l c c r}
\hline \hline 
 & & $J^{\pi}$ &  \\ \cline{2-4}
$\gamma$-decay width (meV)  \hspace{1cm} & $2^+$ & $3^+$ & $1^-$ \\  \hline 
$\avg{\Gamma_{\gamma,\rm sim}}$ & 165.5 & 157.5 & 191.2 \\  
$\avg{\Gamma_{\gamma,\rm exp}}$ & 206 (31) & 240 (58) & 670 (225)  \\  
 & & $J^{\pi}$ &  \\ \cline{2-4}
$\gamma$-decay width (meV) \hspace{1cm} & $2^-$ & $3^-$ & $4^-$ \\ \hline
$\avg{\Gamma_{\gamma,\rm sim}}$ & 172.8 & 169.2 & 153.8 \\ 
$\avg{\Gamma_{\gamma,\rm exp}}$ & 374 (115)  & 404 (100) & 361 (106) \\ \hline \hline
\end{tabular}

\end{table}

We also find that for all choices of the $\gamma$SF parameters described above, the 
partial width fluctuations follow the PTD, similar to what is shown in Fig.~\ref{dist20}. The results for the various cases are included
in the Supplemental Material.

\section{Sensitivity to deviations from PTD}\label{beyond-ptd}

In Sec.~\ref{partial}, we showed that realistic level-density and
$\gamma$SF parameterizations do not lead to any violation of the PTD
for the partial $\gamma$-decay widths. It is interesting, however, to
find out whether the experimental results may be interpreted as
evidence of PTD violation in some channels.  If this were to be the case, then it
would indicate a problem with the conventional statistical-model approach. The authors of 
Ref.~\cite{Koehler2013} used a $\chi^2$ distribution with
$\nu = 0.5$ degrees of freedom instead of the PTD but could not obtain
agreement with the data. However, when the PTD is violated, the partial width 
distribution is not described well by a $\chi^2$ distribution in $\nu$ degrees of freedom. In this section, 
we examine the effect of a realistic PTD violation on the simulated
total width distribution.

\begin{figure}[h!]
\includegraphics[width=0.48\textwidth]{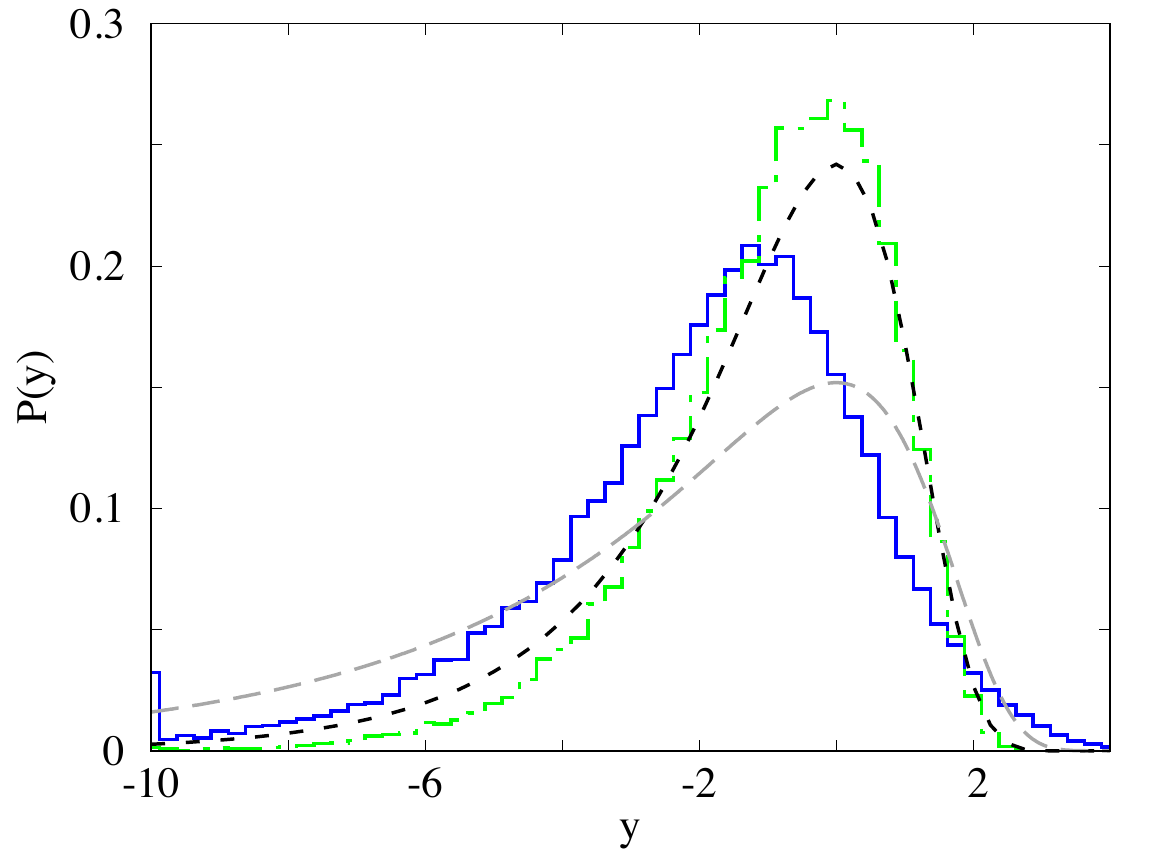}
\caption{\label{ptd-viol} Distribution of $y_c= \ln x_c$ for the model
  of Eq.~(\ref{one-shift}), where $x_c = \hat g_{\mu c}/\avg{\hat
    g_{\mu c}}$ (see Sec.~\ref{partial}). The blue solid
  histogram is $c=1$ (the neutron channel), and the green dashed-dotted histogram is $c=2$ (a $\gamma$ channel).
  The short-dashed black line is the PTD, and the long-dashed grey
  line corresponds to a $\chi^2$ distribution for $x_c$ with $\nu =
  0.5$ degrees of freedom.}
\end{figure}

We obtain a partial width distribution that deviates from the PTD using the
model of Ref.~\cite{Volya2015}.  In this model, the effective
Hamiltonian of Eq.~(\ref{heff-canonical}) is replaced by
\be\label{one-shift} 
H^{\rm eff}_{\mu\nu} = H^{\rm GOE}_{\mu \nu} + Z
\delta_{\mu 1} \delta_{\nu 1} \;.
\ee
 To obtain a large PTD violation, we
use a relatively large imaginary value $Z/\lambda = -0.8 \, i$, as was done in
Ref.~\cite{Volya2015}.  We then examine the distributions of the
quantities $\hat g_{\mu 1}$ and $\hat g_{\mu 2}$, i.e., the
normalized squared projections of the eigenvectors $| \mu \rangle$ of $H^{\rm eff}$ 
onto the first and second basis vectors [see
Eq.~(\ref{sq-proj})]. Following the approach described in
Sec.~\ref{model}, we can identify the basis vectors $\ket{1}$ with the
neutron channel and $\ket{2}$ with a $\gamma$ channel. As before, we
include only the middle half of the GOE spectrum to avoid edge
effects. It was shown in Ref.~\cite{Volya2015} that the distribution
of partial neutron widths $\hat g_{\mu 1}$ is substantially
different from the PTD in this case.  Interestingly, we find that the
distribution of the width $\hat g_{\mu 2}$ for a $\gamma$ channel
is also significantly modified.  Thus, sufficiently large
non-statistical terms in the effective Hamiltonian can cause a
`cross-channel' effect.  The resulting distributions of the
logarithms of the normalized squared projections are shown in
Fig.~\ref{ptd-viol}, along with the PTD.  The figure makes it clear
that neither of the modified distributions is well-described by a $\chi^2$
distribution.

\begin{figure}[h!]
\includegraphics[width=0.48\textwidth]{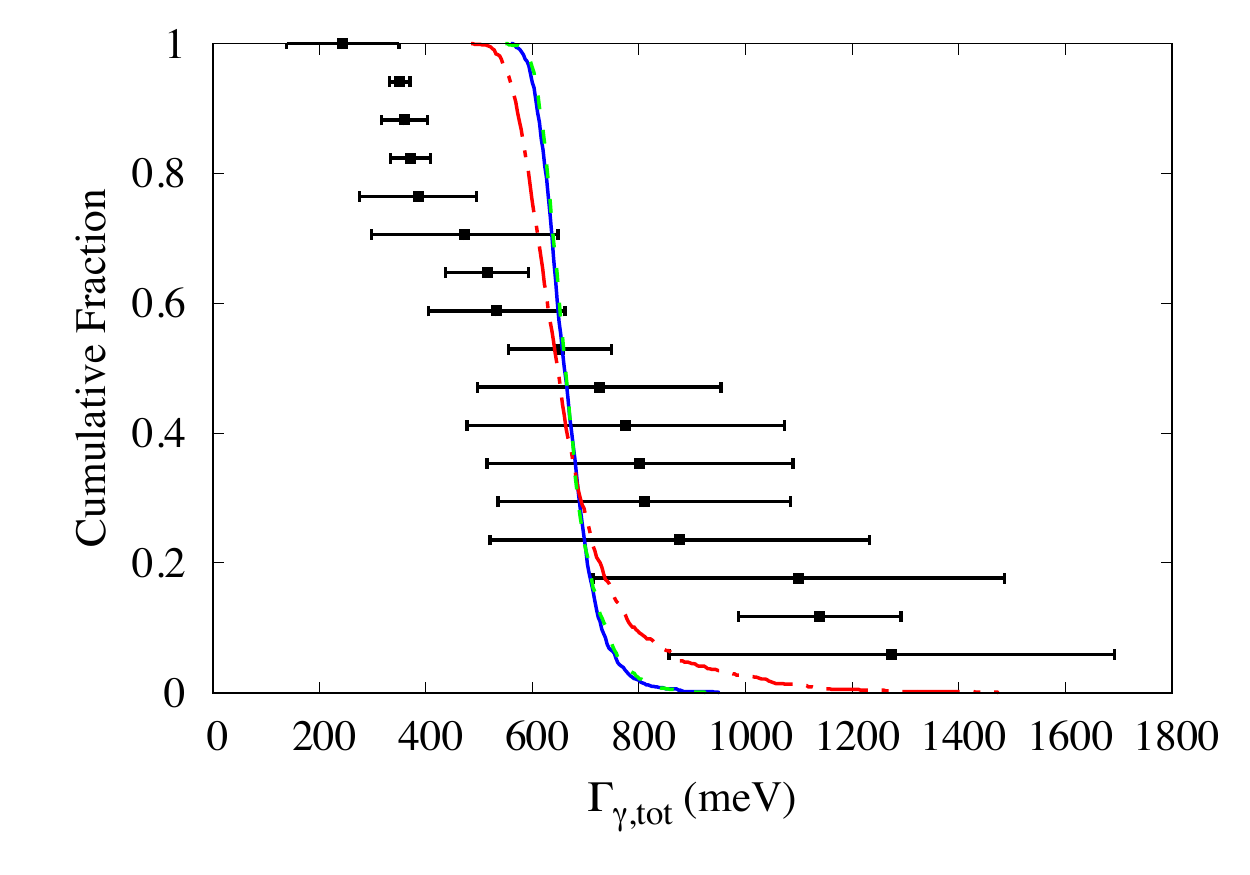}
\caption{\label{total-ptdviol} The cumulative fraction of simulated total $\gamma$-decay width distributions
  compared with data for resonances of spin-parity $1^-$. The blue
  solid line is obtained using PTD fluctuations of the partial
  widths. The red dashed line is obtained using the modified
  distribution corresponding to $c=1$ shown in Fig.~\ref{ptd-viol} for
  the partial-width fluctuations. The green dashed-dotted line is
  obtained using the modified distribution corresponding to $c=2$
  shown in Fig.~\ref{ptd-viol} for the partial-width fluctuations.
  The black dots show the experimental data. The simulated results
  are normalized to match the experimental average total width.}
\end{figure}

We use these modified distributions to generate partial
width fluctuations in our simulation of the total width distribution
described in Sec.~\ref{vary}. Specifically, we replace the quantity
$x_f^2$ in Eq.~(\ref{sim-partial}) with a number drawn from one of
these modified distributions. In Fig.~\ref{total-ptdviol}, we compare the
simulated total $\gamma$-decay width distributions obtained when the partial width distribution is either the PTD 
or one of the above modified distributions with the experimental data for the
$1^-$ resonances. The modified partial width distributions widen the total $\gamma$-decay width distribution
slightly, but not sufficiently to obtain agreement with the data.
Moreover, variations of the $\gamma$SF parameters in the
case of the modified distributions also do not broaden significantly the
$\gamma$-decay width distributions. Thus, we find no evidence that a
modification of the PTD alone can account for the broader
total $\gamma$-decay width fluctuations that are observed in the experiment.

The above conclusion is not unexpected for the following reason. The total
$\gamma$-decay width is the sum of independently distributed random variables,
i.e., the partial width fluctuations $x_f^2$ of
Eq.~(\ref{sim-partial}), each weighted by the appropriate average
partial width.  If a sufficiently large number of final states
contribute roughly equally to the total width, then the central-limit
theorem guarantees that the total $\gamma$-decay width distribution will be a
Gaussian with a very narrow variance. This conclusion holds
regardless of the underlying distribution of the partial widths, 
provided that the distribution does not violate the assumptions of the theorem.
As is evident from our results, partial width distributions that are derived in the framework of the 
the statistical model are consistent with the central-limit theorem.

The total $\gamma$-decay width distribution can be broad only if there exist a small
number of $\gamma$ channels coupled strongly enough to overcome the
restriction of the central-limit theorem. This idea is consistent with
the `doorway' model of Koehler {\em et al.}~in
Ref.~\cite{Koehler2013}, in which the strengths of transitions to all
low-lying final states were multiplied by a factor of $25$. The results of this
model were in good agreement with the experimental data. However, such
a drastic increase in the $\gamma$ transition strength to low-lying
states is outside the relatively large range of conventional $\gamma$SF
models that we have explored above and thus demands a physical
explanation.  In particular, such an enhancement would constitute a
violation of the generalized Brink-Axel hypothesis~\cite{Martin2017}, which states that the strength of a $\gamma$
transition is independent of the details of the initial and final
states at low excitation energies.  Recently, an experiment measured
the photo-absorption strength for $1^{-}$ states of the $^{96}$Mo CN
and found agreement with $\gamma$ decay experiments~\cite{Martin2017}.
This indicates that the generalized Brink-Axel hypothesis holds to a fairly
good approximation in this nucleus and casts doubt on the existence of
the sort of enhancement discussed above.

\section{Conclusion}\label{conclusion}

We have presented a model that is based on a statistical
description of the CN but takes into account the many $\gamma$ channels
coupled to the CN in a semi-realistic way. We applied the model
to the $^{95}$Mo$(n,\gamma)^{96}$Mo* reaction. Using empirical
parameterizations for the level density and $\gamma$SF, we found
that the PTD provides an excellent description of the partial widths
for both the neutron and the $\gamma$-decay channels, in agreement with
the traditional prediction of the statistical model. This result holds for all spin-parity
 values of the CN resonances. We conclude that the net effect of
the large number of $\gamma$-decay channels does not perturb the GOE
statistics of the CN and cannot explain the experimental results of
Ref.~\cite{Koehler2013}. Although it is usually assumed that the
$\gamma$-decay channels have little effect on the GOE statistics of the 
resonances, this has not previously been demonstrated within a
realistic model.

Furthermore, we find that the width of the total $\gamma$-decay width distribution
 is insensitive to large parameter variations of the $E1$ $\gamma$SF.  In particular, 
 the measured width of the distribution of total $\gamma$-decay widths cannot be
  reproduced. 
 We also find that
deviations of the partial-width distributions from PTD (which can in
principle occur for sufficiently strong coupling of the neutron
channel) do not significantly broaden the total $\gamma$-decay width
distributions. This finding follows from the central-limit theorem
and the fact that, for common parameterizations of the level density
and $\gamma$SF, many $\gamma$ channels contribute similarly to the total
width.

The only way to overcome the limitation of the central-limit theorem
is to dramatically increase the $\gamma$ transition strength to a small
group of channels, as investigated in Ref.~\cite{Koehler2013}.  However, such
an enhancement would violate the generalized
Brink-Axel hypothesis and consequently contradict recent experimental
results~\cite{Martin2017}. 

In conclusion, our analysis shows that the results of
Ref.~\cite{Koehler2013} cannot be explained within the
statistical-model framework. Given the fundamental importance of the
statistical model for nuclear-reaction modeling, this discrepancy
should motivate further experimental investigations, both to verify
the findings of Ref.~\cite{Koehler2013} and to test the GOE description of the compound nucleus.\\

\section*{Acknowledgements}
 We thank M. Krti\v{c}ka and P.~E. Koehler for useful discussions, and
A. Richter for making us aware of the experimental results of
Ref.~\cite{Martin2017}. We also thank P.~E. Koehler for providing the
experimental data used here.
This work was supported in part by the U.S.~DOE Grant
 Nos.~DE-FG02-91ER40608 and DE-SC0019521, and by the U.S.~DOE NNSA Stewardship Science
Graduate Fellowship under cooperative agreement No.~DE-NA0003864. The
initial part of this work was performed at the Aspen Center for
Physics, which is supported by a National Science Foundation Grant
No.~PHY-1607611.

\end{document}